\PassOptionsToPackage{pdfpagelabels=false}{hyperref}

\makeatletter
    \newcommand{\dontusepackage}[2][]{%
        \@namedef{ver@#2.sty}{9999/12/31}
        \@namedef{opt@#2.sty}{#1}
    }
\makeatother

\dontusepackage{fixltx2e}

\documentclass[fleqn,usenatbib]{mnras}

\usepackage{amsmath,amssymb}  
\usepackage{booktabs}  
\usepackage{datetime}  
\usepackage[T1]{fontenc}  
\usepackage{graphicx}  
\usepackage{microtype}  
\usepackage{multirow}  
\usepackage{newtxtext}  
\usepackage[slantedGreek]{newtxmath}  
\usepackage{physics}  
\usepackage{siunitx}  
\usepackage{textgreek}  
\usepackage{xcolor}  

\graphicspath{{graphics/}}

\definecolor{MNRASPurple}{HTML}{BC3B9C}

\newdateformat{yearmonthdate}{%
    \THEYEAR\ \monthname[\THEMONTH] \THEDAY
}
\yearmonthdate

\DeclareSIUnit\parsec{pc}
\DeclareSIUnit\h{\text{$h$}}

\hypersetup{
    pdfauthor={Mike (Shengbo) Wang, Florian Beutler, David Bacon},
    pdftitle={Impact of relativistic effects on the primordial non-Gaussianity signature in the large-scale clustering of quasars},
    pdfsubject={cosmology},
    pdfkeywords={cosmology: observations; cosmological parameters; large-scale structure of Universe}
}

\let\oldeqref\eqref
\makeatletter
    \RenewDocumentCommand\eqref{s m}{%
        \IfBooleanTF#1%
        {\textup{\tagform@{\ref*{#2}}}}%
        {\oldeqref{#2}}%
    }
\makeatother

\makeatletter
\patchcmd{\NAT@citex}
    {\@citea\NAT@hyper@{%
        \NAT@nmfmt{\NAT@nm}%
        \hyper@natlinkbreak{\NAT@aysep\NAT@spacechar}%
            {\@citeb\@extra@b@citeb}%
        \NAT@date%
    }}
    {\@citea\NAT@nmfmt{\NAT@nm}\NAT@aysep\NAT@spacechar\NAT@hyper@{\NAT@date}}{}{}
\patchcmd{\NAT@citex}
    {\@citea\NAT@hyper@{%
        \NAT@nmfmt{\NAT@nm}%
        \hyper@natlinkbreak{\NAT@spacechar\NAT@@open\if*#1*\else#1\NAT@spacechar\fi}%
            {\@citeb\@extra@b@citeb}%
        \NAT@date%
    }}
    {\@citea\NAT@nmfmt{\NAT@nm}%
        \NAT@spacechar\NAT@@open\if*#1*\else#1\NAT@spacechar\fi\NAT@hyper@{\NAT@date}%
    }{}{}
\makeatother

\makeatletter
    \newcommand{\linktarget}[1]{\Hy@raisedlink{\hypertarget{#1}{}}}
\makeatother

\newcommand{\affil}[2]{%
    \textsuperscript{\hyperlink{#1}{#2}}%
}
\newcommand{\correspondence}[1]{%
    \thanks{%
        \hspace{-0.5em} Email: \href{mailto:#1}{#1}
    }
}
\newcommand{\orcid}[1]{%
    \hspace{0.35\baselineskip}%
    \raisebox{-1pt}{\href{https://orcid.org/#1}{
        \includegraphics[height=0.7\baselineskip,keepaspectratio]{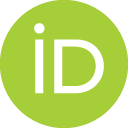}
    }}%
    \hspace{-0.4ex}%
}

\renewcommand{\jcap}{J. Cosmol. Astropart. Phys.}
\newcommand{\jhep}{J. High Energy Phys.}
\newcommand{\is}{iz}

\newcommand{\e}{\kern0.25pt\mathrm{e}\kern0.25pt}
\newcommand{\im}{\kern0.25pt\mathrm{i}\kern0.25pt}

\newcommand{\est}[1]{\hat{#1}}
\newcommand{\measurement}[3]{${#1}^{+#2}_{-#3}$}
\newcommand{\below}[1]{<\!{#1}}

\newcommand{\prob}{\mathbb{P}}
\newcommand{\likelihood}{\mathcal{L}}
\newcommand{\variance}{\operatorname{Var}}

\newcommand{\cH}{\mathcal{H}}
\newcommand{\fNL}{f_\textrm{NL}}

\newcommand{\den}{\delta}
\newcommand{\phitheta}{\phi_{\kern-0.5pt\theta}}
\newcommand{\nbar}{\bar{n}}
\newcommand{\mbar}{\bar{m}}

\newcommand{\los}{\vu*{n}}
\newcommand{\vk}{\vb*{k}}
\newcommand{\vr}{\vb*{r}}

\renewcommand{\vv}{\vb*{v}}

\newcommand{\vur}{\vu*{r}}

\newcommand{\matter}{\textrm{m}}

\newcommand{\evol}{\textrm{e}}
\newcommand{\pivot}{\textrm{p}}
\newcommand{\low}{\textrm{l}}
\newcommand{\high}{\textrm{h}}

\newcommand{\change}[1]{#1}  

\title[Relativistic impact on PNG signature]{%
    Impact of relativistic effects on the primordial non-Gaussianity signature in the large-scale clustering of quasars 
}
\author[M. S. Wang et al.]{%
    \parbox[b]{\linewidth}{
        Mike (Shengbo) Wang%
        \orcid{0000-0002-2652-4043},\affil{affil1}{1}\correspondence{mike.wang@port.ac.uk} %
        Florian Beutler%
        \orcid{0000-0003-0467-5438}\affil{affil1}{2}\affil{affil1}{,1} %
        and David Bacon%
        \orcid{0000-0002-2562-8537}\affil{affil1}{1}%
    }
    \vspace{6pt} \\ \hspace*{-3.5pt}
    \parbox[b]{\linewidth}{
        \linktarget{affil1}{\textsuperscript{\textup{1}}}Institute of Cosmology and Gravitation, University of Portsmouth, Burnaby Road, Portsmouth PO1~3FX, UK\\
        \linktarget{affil2}{\textsuperscript{\textup{2}}}Institute for Astronomy, University of Edinburgh, Royal Observatory, Blackford Hill, Edinburgh EH9~3HJ, UK
    }
}
\date{Accepted 2020 September 23. Received 2020 September 7; in original form 2020 July 6}
\pubyear{202y}
\volume{vvv}

\begin{document}
\label{firstpage}

\pagerange{\pageref{firstpage}--\pageref{lastpage}}
\maketitle

\begin{abstract}
    Relativistic effects in clustering observations have been shown to introduce scale-dependent corrections to the galaxy overdensity field on large scales, which may hamper the detection of primordial non-Gaussianity $\fNL$ through the scale-dependent halo bias. The amplitude of relativistic corrections depends not only on the cosmological background expansion, but also on the redshift evolution and sensitivity to the luminosity threshold of the tracer population being examined, as parametr{\is}ed by the evolution bias $b_\evol$ and magnification bias $s$. In this work, we propagate luminosity function measurements from the extended Baryon Oscillation Spectroscopic Survey (eBOSS) to $b_\evol$ and $s$ for the quasar (QSO) sample, and thereby derive constraints on relativistic corrections to its power spectrum multipoles. Although one could mitigate the impact on the $\fNL$ signature by adjusting the redshift range or the luminosity threshold of the tracer sample being considered, we suggest that, for future surveys probing large cosmic volumes, relativistic corrections should be forward modelled from the tracer luminosity function including its uncertainties. This will be important to quasar clustering measurements on scales $k \sim \SI{e-3}{\h\per\mega\parsec}$ in upcoming surveys such as the Dark Energy Spectroscopic Instrument (DESI), where relativistic corrections can overwhelm the expected $\fNL$ signature at low redshifts $z \lesssim 1$ and become comparable to $\fNL \simeq 1$ in the power spectrum quadrupole at redshifts $z \gtrsim 2.5$.
\end{abstract}


\raggedbottom

\section{\texorpdfstring{\textls{Introduction}}{Introduction}}
\label{sec:introduction}

It is known that primordial non-Gaussianity~(PNG), which encodes dynamics of the inflationary period in the early Universe, leaves an imprint in the large-scale structure~(LSS) at late times not only in higher-order statistics such as the bispectrum, but also in the clustering of virial{\is}ed haloes by introducing a scale-dependent modification to the large-scale tracer bias~\citep{Dalal_2008,Matarrese_2008,Slosar_2008}. For the local type of PNG~$\fNL$, although the strongest constraint yet comes from observations of the cosmic microwave background~(CMB) by \textit{Planck}\footnote{\href{http://www.esa.int/planck}{\texttt{esa.int/planck}}}~\citep[$\fNL = 0.9\pm5.1$; Planck Collaboration,][]{Planck_2019}, upcoming LSS probes such as the Dark Energy Spectroscopic Instrument\footnote{\href{https://desi.lbl.gov}{\texttt{desi.lbl.gov}}}~(DESI) and \textit{Euclid}\footnote{\href{https://www.euclid-ec.org/}{\texttt{euclid-ec.org}}} are forecast to offer competitive constraints with uncertainties of $\order{1}$~\citep{Font_Ribera_2014,Amendola_2018,Mueller_2018}, with current galaxy surveys such as the extended Baryon Oscillation Spectroscopic Survey\footnote{\href{https://www.sdss.org/surveys/eboss/}{\texttt{sdss.org/surveys/eboss}}}~(eBOSS) already achieving uncertainties of $\order{10}$~\citep{Castorina_2019}.

Despite the relativistic nature of gravitational theories governing structure formation, the Newtonian description of fluctuations in the distribution of galaxies is usually adequate as relativistic effects are suppressed below the Hubble horizon scale. In the past, the modelling of fully relativistic galaxy clustering has been unnecessary to obtain cosmological parameter constraints, as cosmic variance dominates over any corrections. With the next generation of galaxy surveys probing far wider and deeper cosmic volumes, however, such approximate prescriptions might no longer be sufficient to attain unbiased constraints. The necessary relativistic corrections for galaxy clustering observations have been derived by \citet{Yoo_2009}, \citet{Bonvin_2011} and \citet{Challinor_2011}. Many subsequent works have demonstrated their importance for constraining cosmological parameters, in particular $\fNL$, as its scale-dependent signature on large scales can be disguised as relativistic effects~\citep{Bruni_2012,Jeong_2012,Bertacca_2012,Camera_2015,Alonso_2015,Fonseca_2015,Raccanelli_2016a,Raccanelli_2016b,Lorenz_2018}. On the other hand, the investigation of relativistic corrections in itself is a valuable exercise, as it offers tests of relativistic gravitational theories on cosmological scales~\citep{Lombriser_2013,Bonvin_2014}, including the equivalence principle~\citep{Bonvin_2020}. Future galaxy surveys like DESI are forecast to deliver the first detections of these relativistic corrections~\citep{Beutler_2020}.

One crucial aspect of relativistic corrections is that their total amplitude does not only depend on the cosmological and gravitational models, but also on the background number density of the tracer population being examined through its redshift evolution and sensitivity to the luminosity threshold of observations, as respectively captured by parameters known as the evolution bias~$b_\evol$ and magnification bias~$s$. Previous works have mostly only considered relativistic effects in Fisher forecasts for $\fNL$ by assuming fiducial values of $b_\evol$~and~$s$, but the exact dependence of these parameters on redshift and the luminosity threshold, as well as how their uncertainties propagate to the observed power spectrum, remains much less clear. In this work, we concret{\is}e these considerations for quasars~(QSO), which are an ideal tracer for detecting $\fNL$ thanks to their high redshift range and bias, and proceed as follows:
\begin{enumerate}
    \item We first review in section~\ref{sec:relativistic clustering} general relativistic corrections in galaxy clustering to linear order, including contributions from evolution bias~$b_\evol$ and magnification bias~$s$ which we shall formally introduce. This motivates the need for determining the tracer luminosity function;
    \item Based on the previous work by~\citet{Palanque_Delabrouille_2016}, we fit the quasar luminosity function with eBOSS QSO measurements in section~\ref{sec:luminosity function}, before deriving constraints on $b_\evol$, $s$ and thus relativistic corrections in section~\ref{sec:relativistic constraints};
    \item In section~\ref{sec:scale-dependent modifications}, we compare scale-dependent modifications to the quasar power spectrum due to relativistic corrections and due to $\fNL$ at different redshifts for two different magnitude thresholds, and discuss in section~\ref{sec:conclusions} the need to include luminosity function constraints in forward modelling of relativistic clustering statistics for future galaxy surveys.
\end{enumerate}

\section{\texorpdfstring{\textls{Relativistic Clustering of Galaxies}}{Relativistic Clustering of Galaxies}}
\label{sec:relativistic clustering}

Whilst the Newtonian description of galaxy clustering is appropriate for observations on sub-horizon scales, as the clustering scale~$k^{-1}$ approaches the horizon scale~$\cH^{-1}$, where $\cH(z)$ is the conformal Hubble parameter at redshift~$z$, \change{the observed galaxy overdensity field~$\den$ receives relativistic corrections of~$\order{\cH/k}$ or higher that are otherwise suppressed},
\change{
    \begin{alignat}{3}
        \den(\vr, z) = {} &&& b_1 \den_\matter - \frac{1}{\cH} \vur \vdot \partial_r \vv \nonumber \\
        &&& - g_1(z) \vur \vdot \vv - (b_\evol - 3) \cH \nabla^{-2} \grad \vdot \vb*{v} \nonumber \\
        &&& + \frac{1}{\cH} \Phi' - (2 - 5s) \Phi + \Psi + g_1(z) \Psi + \dotsb \,.
        \label{eq:relativistic clustering}
    \end{alignat}
}%
\change{Here $b_1(z)$ is the scale-independent tracer bias\footnote{\change{We will later consider scale-dependent modifications in section~\ref{sec:scale-dependent modifications}.}} with respect to the matter density contrast~$\den_\matter$ in the comoving synchronous gauge, $\vv$ is the peculiar velocity in the Newtonian gauge, $\Phi$~and~$\Psi$ are the Bardeen potentials, $g_1(z)$ is a dimensionless quantity given by
    \begin{equation}
        g_1(z) = \frac{\cH'}{\cH^2} + \frac{2-5s}{\cH\chi} + 5s - b_\evol \,,
    \end{equation}
$\chi(z)$ is the comoving distance, and $'$ denotes a conformal time derivative~\citep{Bonvin_2011,Challinor_2011}.} The quantities $b_\evol$~and~$s$ are the \emph{evolution} and \emph{magnification biases}, which do not a priori follow from a background cosmological model but are rather derived at a given redshift from
\begin{subequations}
    \label{eq:relativistic biases}
    \begin{align}
        & b_\evol(z) = - \pdv{\ln\nbar(z; \below{\mbar})}{\ln(1 + z)} \,, \\
        & s(z) = \eval{\pdv{m}}_{\mbar} {\lg\nbar(z; \below{m})}
    \end{align}
\end{subequations}
with $\lg \equiv \log_{10}$, where $\nbar(z; \below{m})$ is the underlying comoving number density of the tracer population below a given absolute magnitude~$m$, and $\bar{m}$ is the absolute magnitude threshold of the observed tracer sample~\citep{Challinor_2011}.

\change{In equation~\eqref{eq:relativistic clustering}, we have neglected lensing magnification, time delay and the integrated Sachs--Wolfe~(ISW) effect, which are integrated terms involving the Bardeen potentials and cannot be easily included in a Cartesian power spectrum model. All of these terms may affect cosmological parameter inference, as shown by recent studies of their relative importance using the angular power spectrum or correlation function~\citep{Namikawa_2011,Raccanelli_2016,Raccanelli_2016a,Lorenz_2018,Jelic_Cizmek_2020}. In this work, we shall instead focus on the Doppler terms involving the peculiar velocity and the local potential terms only, and consider their scale-dependent signature in the plane-parallel limit where $\mu \equiv \hat{\vk} \vdot \vur = \hat{\vk} \vdot \los$ does not vary for a fixed global line of sight~$\los$. Using the linear{\is}ed Einstein equations for a \textLambda{CDM} universe,
\begin{subequations}
    \begin{align}
        & \vb*{v} = - \im \frac{\cH}{k} f \den_\matter \vu*{k} \,, \\
        & \Phi = - \frac{3}{2} \qty(\frac{\cH}{k})^2 \den_\matter \,, \\
        & \cH^{-1} \Phi' = \qty(\frac{\cH'}{\cH^2} - 1) \qty(\frac{\cH}{k})^2 f \den_\matter - \Phi \,,
    \end{align}
\end{subequations}
where $f(z)$ is the linear growth rate and $\Phi = \Psi$ in the absence of anisotropic stress~\citep{Bruni_2012,Jeong_2012,Bertacca_2012}, we can recast equation~\eqref{eq:relativistic clustering} as
    \begin{equation}
        \den(\vk, z) = \qty[b_1 + f \mu^2 + \im \frac{\cH}{k} g_1(z) f \mu + \qty(\frac{\cH}{k})^2 g_2(z)] \den_\matter(\vk, z) \,,
        \label{eq:relativistic corrections}
    \end{equation}
where we have introduced a second dimensionless quantity
    \begin{equation}
        g_2(z) \equiv - \qty(b_\evol - 3) f + \qty(\frac{\cH'}{\cH^2} - 1) \qty\big[g_1(z) + f - (2 - 5s)] \,.
    \end{equation}
By employing the Friedman equations,\footnote{We neglect radiation and spatial curvature.} we can rewrite
    \begin{equation}
        \frac{\cH'}{\cH^2} = 1 - \frac{3}{2} \Omega_\matter
    \end{equation}
in terms of the matter density parameter~$\Omega_\matter(z)$. The quantities parametr{\is}ing relativistic corrections are thus
\begin{subequations}
    \label{eq:relativistic contributions}
    \begin{align}
        & g_1(z) = \qty(3 - b_\evol - \frac{3}{2} \Omega_\matter) - (2 - 5s) \qty(1 - \frac{1}{\cH\chi}) \,, \\
        & g_2(z) = \qty(3 - b_\evol - \frac{3}{2} \Omega_\matter) f - \frac{3}{2} \Omega_\matter \qty\big[g_1(z) - (2 - 5s)] \,,
    \end{align}
\end{subequations}
and they depend not only on the cosmological density parameters through the accelerating background expansion, but also on the tracer sample in question through its evolution and magnification biases.}

Therefore to determine the relativistic corrections in equation~\eqref{eq:relativistic clustering} or \eqref{eq:relativistic corrections}, two ingredients are needed: (1) a background cosmological model; (2) the tracer \emph{luminosity function}~(LF) $\phi(m, z)$ from which the underlying comoving number density
    \begin{equation}
        \nbar(z; \below{\mbar}) = \int_{-\infty}^{\mbar} \dd{m} \phi(m, z)
        \label{eq:tracer number density}
    \end{equation}
below the absolute magnitude threshold~$\mbar$ can be derived -- this is our focus in the next section.

\section{\texorpdfstring{\textls{Quasar Luminosity Function}}{Quasar Luminosity Function}}
\label{sec:luminosity function}

Determining the tracer luminosity function is not only important for modelling relativistic corrections, it could also be a significant source of uncertainty for constraining PNG. In this work, we examine quasars as a single tracer for detecting $\fNL$ thanks to their high tracer bias and redshift range. We attempt to recover their evolution and magnification biases from their luminosity function, before propagating these measurements to relativistic corrections to the power spectrum multipoles.

To this end, we consider eBOSS QSO LF measurements obtained by \citet[Table~A.1 therein]{Palanque_Delabrouille_2016} for the redshift range~$0.7 < z < 4$, which are corrected for observational systematics such as completeness and bandpass redshifting of spectra (i.e. $K$-correction). We describe the empirical quasar luminosity function with the \emph{pure luminosity evolution}~(PLE) model~\citep{Boyle_2000,Richards_2006,Palanque_Delabrouille_2016},
    \begin{equation}
        \phi(m, z) = \frac{\phi_\ast}{10^{0.4 (\alpha + 1) [m - m_\ast(z)]} + 10^{0.4 (\beta + 1) [m - m_\ast(z)]}} \,,
        \label{eq:PLE LF model}
    \end{equation}
which is a double power law with bright- and faint-end indices~$\alpha$~and~$\beta$ that may differ depending on the redshift~$z$ relative to the pivot redshift~$z_\pivot = 2.2$. Here $\phi_\ast$ is the overall normal{\is}ation constant, and
    \begin{equation}
        m_\ast(z) = m_\ast(z_\pivot) - \frac{5}{2} \qty[k_1 (z - z_\pivot) + k_2 (z - z_\pivot)^2]
        \label{eq:characteristic magnitude}
    \end{equation}
is the characteristic absolute magnitude, where $k_1$~and~$k_2$ are redshift evolution parameters that can also differ between low redshift~$z < z_\pivot$ and high redshift~$z > z_\pivot$. Therefore this is a parametric model with 10~parameters, $\theta = \qty{\phi_\ast, m_\ast(z_\pivot), \alpha_\low, \beta_\low, k_{1\low}, k_{2\low}, \alpha_\high, \beta_\high, k_{1\high}, k_{2\high}}$, where subscripts `l'~and~`h' denote `low redshift' and `high redshift' respectively.

\subsection{Likelihood function}

Without re-performing the iterative luminosity function fitting procedure on the raw quasar count data in \citet{Palanque_Delabrouille_2016}, we adopt the likelihood inference approach outlined in \citet{Pozzetti_2016} for simplicity. For absolute magnitude and redshift bins $(m_i, z_i)$ indexed by $i$, the quasar number count $\est{N}_i$ follows the \change{Poisson distribution with} logarithmic probability density function~(PDF)
    \begin{equation}
        \ln\prob\qty\big(\est{N}_i \vert N_i) = \est{N}_i \ln{N_i} - {N_i} - \ln\Gamma\qty\big(\est{N}_i)
        \label{eq:number count distribution}
    \end{equation}
with variance $\variance\qty\big(\est{N}_i) = N_i$, where \change{$\Gamma$ denotes the gamma function and} the expected number count is given by
    \begin{equation}
        N_i = \ev{\est{N}_i} = \int_{\text{bin-$i$}} \dd{z} \dv{V(z)}{z} \int_{\text{bin-$i$}} \dd{m} \phitheta(m, z) \,.
    \end{equation}
Here $\phitheta(m, z)$ is the PLE luminosity function~\eqref{eq:PLE LF model} with model parameters~$\theta$ and
    \begin{equation}
        \dd{V(z)} = 4\uppi r^2 \dv{r}{z} \dd{z}
    \end{equation}
is the differential comoving volume, where $r(z)$ is the radial comoving distance.

To obtain an approximate likelihood for the parametric luminosity function model, we first note that the binned luminosity function~$\est{\phi} \propto \est{N}$ and thus the estimated uncertainty on $\ln\est{\phi}$ is $\sigma = {\est{N}}^{-1/2}$. Expanding the PDF~\eqref{eq:number count distribution} around its maximum, we obtain the quadratic form
    \begin{equation}
        \ln\likelihood(\theta) - \ln\likelihood_\textrm{max} \simeq - \frac{1}{2} \sum_i \frac{x_i^2}{\sigma_i^2} \,,
    \end{equation}
where $\sigma_i^2 = \flatfrac{1}{\est{N}_i}$ and
    \begin{equation}
        x_i^2(\theta) = 2 \qty[1 - \frac{\phitheta(m_i, z_i)}{\est{\phi}_i} + \ln\frac{\phitheta(m_i, z_i)}{\est{\phi}_i}] \,.
    \end{equation}
Therefore the likelihood function we shall use to infer the best-fitting luminosity function model is
    \begin{equation}
        \ln\likelihood(\theta) = - \sum_i \frac{1}{\sigma_i^2} \qty[1 - \frac{\phitheta(m_i, z_i)}{\est{\phi}_i} + \ln\frac{\phitheta(m_i, z_i)}{\est{\phi}_i}] \,,
        \label{eq:likelihood function}
    \end{equation}
where we have neglected the additive normal{\is}ation constant.

\subsection{Best-fitting models}
\label{sec:best-fitting models}

By sampling the PLE model parameters from the likelihood function~\eqref{eq:likelihood function} with the Markov chain Monte Carlo~(MCMC) sampler \textsc{zeus}\footnote{\href{https://github.com/minaskar/zeus}{\texttt{github.com/minaskar/zeus}}}~\citep{Karamanis_2020}, we have re-fitted the quasar luminosity function from the eBOSS QSO measurements. Because of the exchange symmetry between the power-law indices $\alpha$~and~$\beta$ in equation~\eqref{eq:PLE LF model}, we have imposed the constraint~$\alpha < \beta$ to avoid a multimodal posterior distribution. The PLE parameters are estimated by the sample posterior medians, as reported in Table~\ref{tab:PLE model fits}, with a reduced chi-square value of $\chi^2\big/\textrm{d.o.f.} = \flatfrac{105}{77} \approx 1.36$ per degree of freedom (d.o.f.).
\begin{table}
    \centering
    \caption{Posterior median estimates of the PLE model parameters (see equation~\ref{eq:PLE LF model}) for the eBOSS QSO LF measurements.}
    \bgroup\setlength{\tabcolsep}{1.2em}
    \begin{tabular}{rrr}
        \toprule[1pt]
        \multirow{2}{*}{Parameter} & \multicolumn{2}{c}{Redshift range} \\[1.25ex]
        & \numrange[range-phrase=--]{0.68}{2.2} & \numrange[range-phrase=--]{2.2}{4.0} \\
        \specialrule{0.75pt}{1pt}{4pt}
        $\lg\phi_\ast$ & \multicolumn{2}{c}{\measurement{-26.20}{0.21}{0.20}} \\[1.5ex]
        $m_\ast(z_\pivot)$ & \multicolumn{2}{c}{\measurement{-5.76}{0.09}{0.08}} \\[1.5ex]
        $\alpha$ & \measurement{-3.27}{0.17}{0.19} & \measurement{-2.57}{0.08}{0.09} \\[1.5ex]
        $\beta$ & \measurement{-1.40}{0.06}{0.06} & \measurement{-1.21}{0.10}{0.09} \\[1.5ex]
        $k_1$ & \measurement{-0.10}{0.08}{0.09} & \measurement{-0.37}{0.09}{0.09} \\[1.5ex]
        $k_2$ & \measurement{-0.40}{0.06}{0.06} & \measurement{-0.01}{0.06}{0.06} \\
        \specialrule{1pt}{4pt}{1pt}
    \end{tabular}
    \egroup
    \label{tab:PLE model fits}
\end{table}

\begin{figure*}
    \centering
    \includegraphics[width=\linewidth]{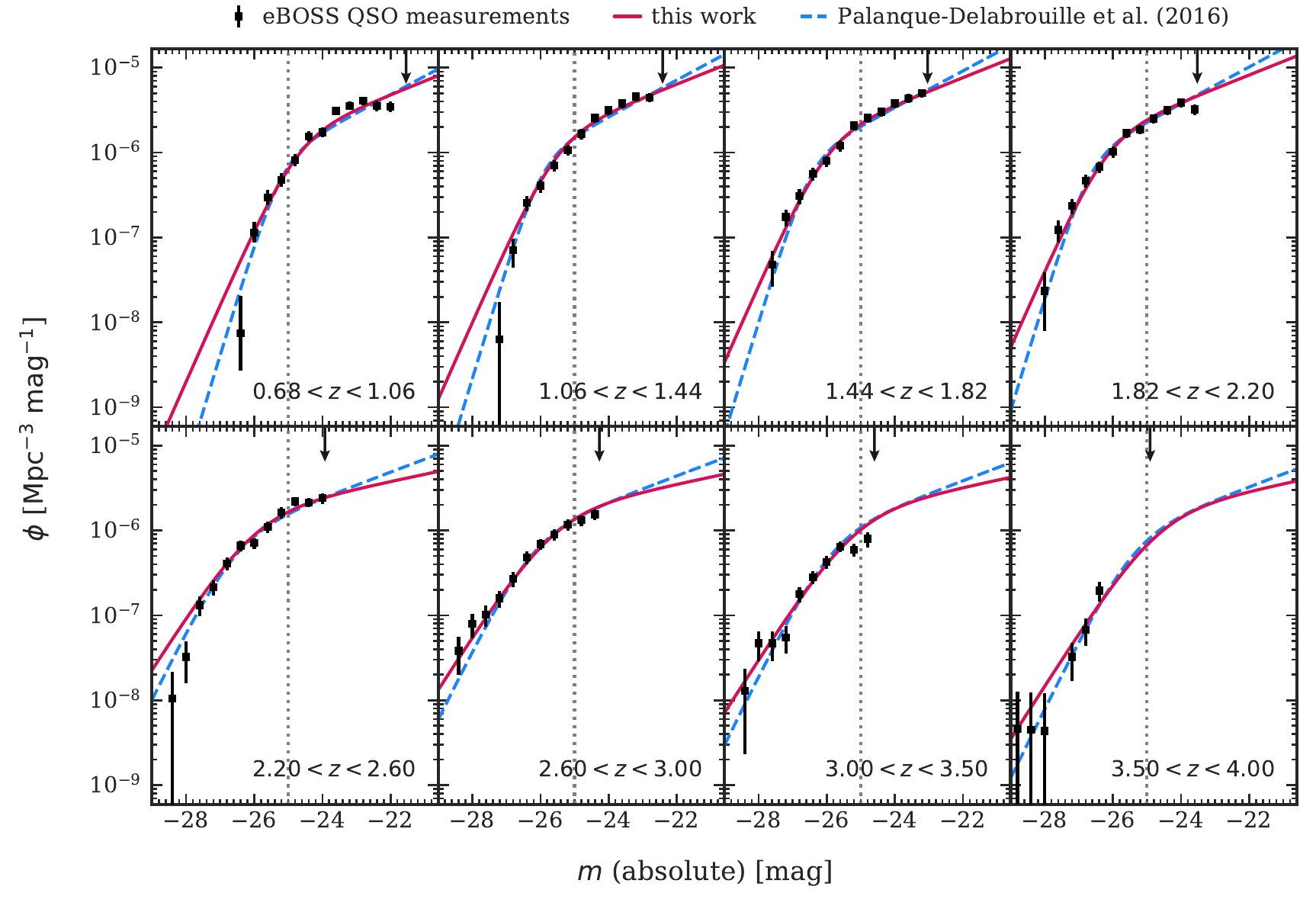} \\[-3pt]
    \caption{Best-fitting quasar luminosity functions under the PLE model~\eqref{eq:PLE LF model} in eBOSS QSO redshift bins. Measurements with uncertainties and the best-fitting model shown in dashed blue lines are taken from \citet{Palanque_Delabrouille_2016}. The best-fitting model of this work (see Table~\ref{tab:PLE model fits}) is shown in solid red lines. The downward pointing arrows mark the limiting absolute magnitude corresponding to the apparent magnitude cut~$g = 22.5$ in each redshift bin. The vertical dotted lines mark the absolute magnitude threshold~$\mbar = -25$ used in this work.}
    \label{fig:luminosity function}
\end{figure*}
We note that there appears to be some discrepancy between our fitted parameters and the results in \citet{Palanque_Delabrouille_2016}, so we compare both best-fitting models with the eBOSS QSO measurements in Fig.~\ref{fig:luminosity function}. In all redshift bins the two fitted models are in reasonable agreement with measurements and are virtually indistinguishable across a wide magnitude range. Noticeable differences only appear either at the very faint end below the limiting absolute magnitude corresponding to the $g$-band apparent magnitude cut $g = 22.5$,\footnote{The $g$-band apparent magnitude is normal{\is}ed to the absolute magnitude $\mbar(z) = g - \mu(z) - [K(z) - K(z=2)]$, where $\mu(z)$ is the distance modulus, $K(z)$ is the $K$-correction, and the normal{\is}ation redshift $z = 2$ is close to the median redshift of the eBOSS QSO sample~\citep{Palanque_Delabrouille_2016}.} which is not constrained by any measurements, or at the very bright end, where uncertainties are comparatively large. We attribute such discrepancies to the fact that \citet{Palanque_Delabrouille_2016} were able to fit the raw quasar number counts corrected for systematics whereas we have only fitted the binned luminosity function reported in their final results.\footnote{\change{It is also worth mentioning that recently \citet{Caditz_2017} noted a possible error in the $K$-correction applied to the eBOSS QSO data sets by~\citet{Palanque_Delabrouille_2016}, which could have an impact on the fitted luminosity function.}} As we shall see in the next section, constraints on the relativistic corrections propagated from these best-fitting luminosity function models are \change{broadly} statistically consistent and have no significant impact on \change{the findings of} our analysis.

\section{\texorpdfstring{\textls{Constraints on Relativistic Corrections}}{Constraints on Relativistic Corrections}}
\label{sec:relativistic constraints}

Having determined the quasar luminosity function, we now proceed to constrain relativistic corrections to quasar clustering statistics by propagating the sampled luminosity function parameters in the form of MCMC chains to evolution and magnification biases. To do so, we specify the Planck15 {\textLambda}CDM cosmology with~$(h, \Omega_{\Lambda,0}, \Omega_{\matter,0}) = (0.6790, 0.6935, 0.3065)$~\citep[Planck Collaboration,][]{Planck_2016}, which is a choice consistent with~\citet{Palanque_Delabrouille_2016}. We also specify a fiducial absolute magnitude threshold $\mbar = -25$ based on the last eBOSS QSO redshift bin.

\subsection{Constraints on relativistic biases}
\label{subsec:relativistic bias constraints}

\begin{figure}
    \centering
    \includegraphics[width=\linewidth]{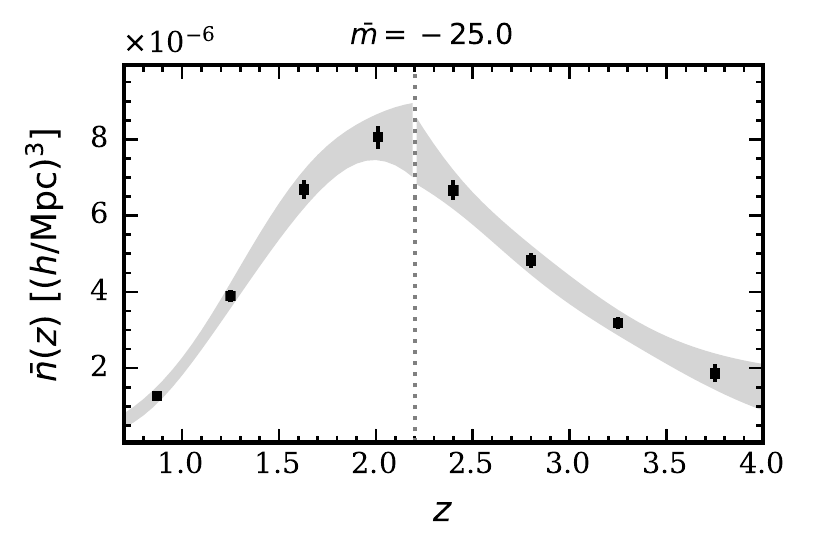} \\[-3pt]
    \caption{Derived measurements of the quasar comoving number density $\nbar(z)$ below the absolute magnitude threshold $\mbar = -25$ from the best-fitting eBOSS QSO LF in this work (see Table~\ref{tab:PLE model fits}). Data points with error bars are measurements within the \SI{68}{\percent} credible interval for the eBOSS QSO redshift bins. The shaded grey regions show the \SI{95}{\percent} credible interval. The vertical dotted line marks the pivot redshift $z_\pivot = 2.2$.}
    \label{fig:tracer number density}
\end{figure}

We first compute the quasar comoving number density~$\nbar(z)$ from equation~\eqref{eq:tracer number density} by numerically integrating our best-fitting luminosity function model~$\phi(m, z)$ up to the absolute magnitude threshold~$\mbar$. In Fig.~\ref{fig:tracer number density}, we show the derived measurements of $\nbar(z)$ from sampled luminosity function parameters within the \SI{95}{\percent}~credible interval across the redshift range~$0.7 < z < 4$; for the eBOSS QSO redshift bins, we also show the measurements of $\nbar(z)$ with error bars corresponding to the \SI{68}{\percent}~credible interval. \change{The small apparent discontinuity in~$\nbar(z)$ corresponds to redshift~$z = z_\pivot$, which divides some subsets of the combined eBOSS QSO data~\citep{Palanque_Delabrouille_2016}. The presence of the pivot redshift~$z_\pivot$ is also a feature of the empirical models currently used for the quasar luminosity function, where the model parameters can suddenly change. This may have possible links to the physics of quasar formation around that epoch in history and/or the fact that the double power-law form assumed for the quasar luminosity function is no longer adequate at higher redshifts~\citep{Caditz_2017,Caditz_2018}}.

Next, we compute the evolution bias~$b_\evol$ and magnification bias~$s$ from equation~\eqref{eq:relativistic biases} by numerical differentiation with redshift step size~$\upDelta z = 0.001$. We have found that, based on the eBOSS QSO LF measurements, $b_\evol$ can be an order of magnitude larger than $s$, although both appear in relativistic corrections at the same orders in equation~\eqref{eq:relativistic contributions}. One interesting comparison one could make for~$b_\evol(z)$ is with the analytic estimate from the universal mass function~(UMF) of haloes, although the validity of this approach is only limited to tracer sample selection that is insensitive to the halo merger history~\citep{Jeong_2012}. The evolution bias predicted from the UMF is given by
    \begin{equation}
        b_\evol(z) = \delta_\textrm{c} f(z) [b_1(z) - 1] \,,
    \end{equation}
where $\delta_\textrm{c} \approx 1.686$ is the critical density of spherical collapse. \change{Here we consider a simple redshift evolution model for the quasar linear bias~$b_1(z) = \flatfrac{1.2}{D(z)}$, whose value increases from~$1.7$ to~$4.7$ almost linearly in the eBOSS QSO redshift range~$0.7 < z < 4$. This bias model is based on the DESI baseline survey~\citep[DESI Collaboration,][]{DESI_2016} and does not account for possible luminosity dependence.} Based on power-law fitting to observed quasar clustering amplitudes, studies have found that the luminosity dependence of quasar bias appears to be rather weak, at least at low and intermediate redshifts possibly because quasars reside in a broad range of haloes of different masses~\citep{White_2012,Shen_2013,Krolewski_2015}. However, some current models and observations hint at a greater level of luminosity dependence at higher redshifts and luminosity ranges, but such quasars are rare and the luminosity dependence of their bias can only be better constrained with larger future data sets~\citep{Shen_2009,Croton_2009,Conroy_2012,Timlin_2018}.

\begin{figure}
    \centering
    \includegraphics[width=\linewidth]{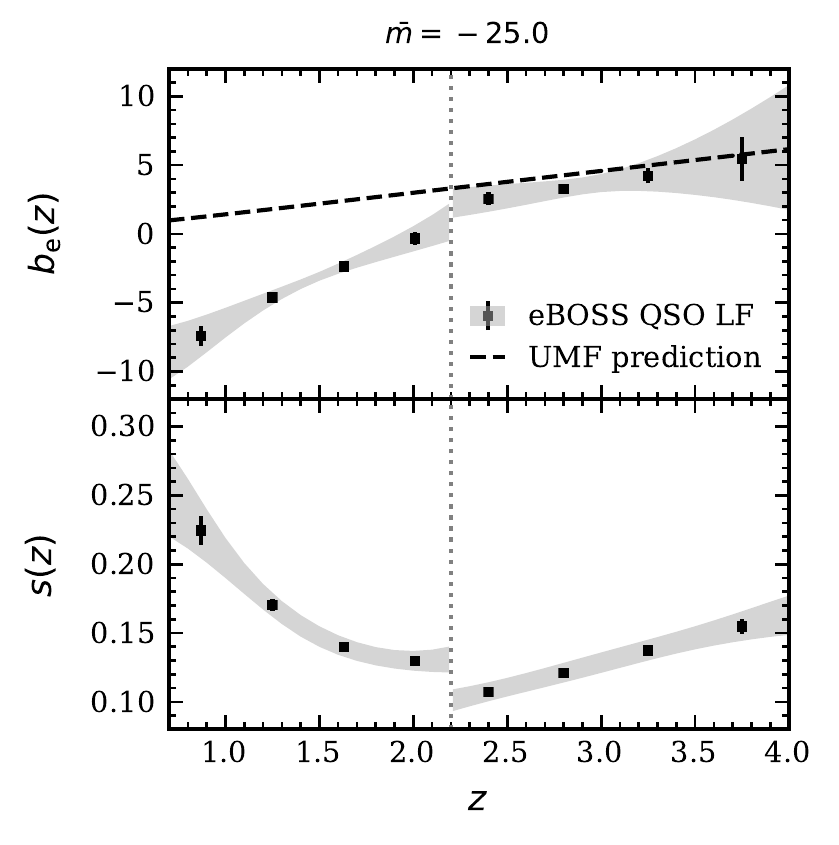} \\[-3pt]
    \caption{Derived measurements of evolution bias~$b_\evol$ and magnification bias~$s$ at the absolute magnitude threshold~$\mbar = -25$ from the best-fitting eBOSS QSO LF in this work (see Table~\ref{tab:PLE model fits}). Data points with error bars are measurements within the \SI{68}{\percent}~credible interval for the eBOSS QSO redshift bins. The shaded grey regions show the \SI{95}{\percent} credible interval. The vertical dotted lines mark the pivot redshift~$z_\pivot = 2.2$. \change{The cause of the discontinuities at $z_\pivot$ in both $b_\evol$~and~$s$ is unclear and could be attributed to unknown systematics in the high-redshift QSO sample~\citep{Kulkarni_2019}.}}
    \label{fig:relativistic biases}
\end{figure}

In Fig.~\ref{fig:relativistic biases}, we show the derived measurements of $b_\evol$~and~$s$ for $0.7 < z < 4$ within the \SI{95}{\percent}~credible interval and their measurements in eBOSS QSO redshift bins with \SI{68}{\percent}~level uncertainties, together with the UMF prediction. \change{Similar to the constraints on comoving number density~$\nbar(z)$, uncertainties of $b_\evol$~and~$s$ at each redshift are derived from samples of their values calculated from MCMC chains of the luminosity function parameters (which may differ on different sides of the pivot redshift~$z_\pivot$).} We note that, although the UMF prediction is in reasonable agreement with our measurements at high redshifts, it does not capture the behaviour of the negative evolution bias values below redshift $z \simeq 2$. This is perhaps unsurprising given the limitation of the UMF prediction and the simplicity of our quasar bias model. \change{As is the case for comoving number density, there is an apparent discontinuity at the pivot redshift $z_\pivot = 2.2$ in both $b_\evol$~and~$s$. However, these discontinuities are now large enough that even the \SI{95}{\percent}~uncertainty bounds are inconsistent across the pivot redshift. Unfortunately, we have checked that this problem still persists with the luminosity functions fitted by \citet{Palanque_Delabrouille_2016} and \citet{Caditz_2017}, so it is not due to our fitting procedure. Although the cause of these discontinuities has been attributed to the form of the empirical luminosity function, the largeness of the discrepancies could indicate unknown systematics in the eBOSS QSO sample at high redshifts, as noted by~\cite{Kulkarni_2019}. Future survey data may hopefully be able to resolve this issue.}

\begin{figure}
    \centering
    \includegraphics[width=\linewidth]{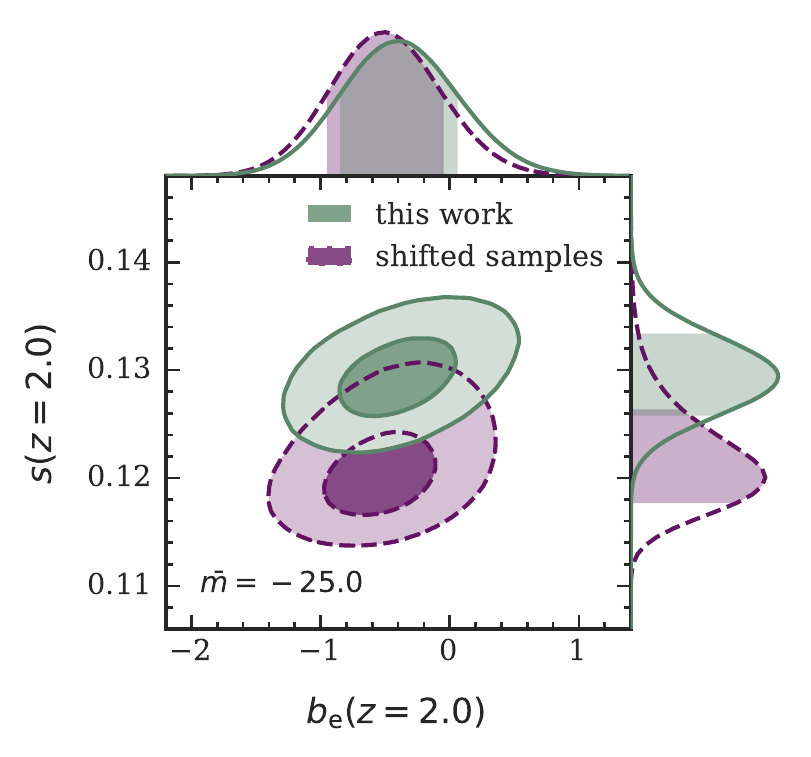} \\[-3pt]
    \caption{Constraints on evolution bias~$b_\evol$ and magnification bias~$s$ at redshift~$z = 2$ and absolute magnitude threshold~$\mbar = -25$ from the eBOSS QSO LF under the PLE model~\eqref{eq:PLE LF model}. The solid green contours show the \SI{68}{\percent} and \SI{95}{\percent} credible regions of the joint posterior distribution sampled from the likelihood function~\eqref{eq:likelihood function} ({this work}). The dashed purple contours are from the same samples except shifted to coincide with the best-fitting PLE model parameters from \citet{Palanque_Delabrouille_2016} ({shifted samples}). The shaded regions in the top and right-hand panels show the \SI{68}{\percent}~credible intervals of the marginal posterior distributions.}
    \label{fig:constraint consistency}
\end{figure}

In section~\ref{sec:best-fitting models}, \change{we have also noted that our best-fitting luminosity function under the PLE model is somewhat discrepant from that of \citet{Palanque_Delabrouille_2016} for the same underlying eBOSS QSO sample (possibly affected by unknown systematics)}, although the parameter estimates have similar uncertainties. To investigate the impact of this on the measured relativistic bias parameters, we shift our sampled luminosity function parameter chains so that the shifted posterior median estimates would coincide precisely with the best-fitting PLE parameters in \citet{Palanque_Delabrouille_2016}, and then we propagate the resultant parameter samples to constraints on $b_\evol$~ and~$s$. In Fig.~\ref{fig:constraint consistency}, we show that the joint $(b_\evol, s)$ constraints at redshift~$z = 2$ from our original parameter samples and the shifted samples are broadly consistent. This is particularly the case for evolution bias~$b_\evol$, which we shall see dominates the relativistic corrections over magnification bias~$s$. We have also checked that the joint $(b_\evol, s)$ constraints from the original and shifted samples are consistent at other redshifts, e.g. $z = 0.87, 3.75$ which are respectively the lowest and highest eBOSS QSO redshift bins.

\subsection{Constraints on the relativistic correction function}

\begin{figure}
    \centering
    \includegraphics[width=\linewidth]{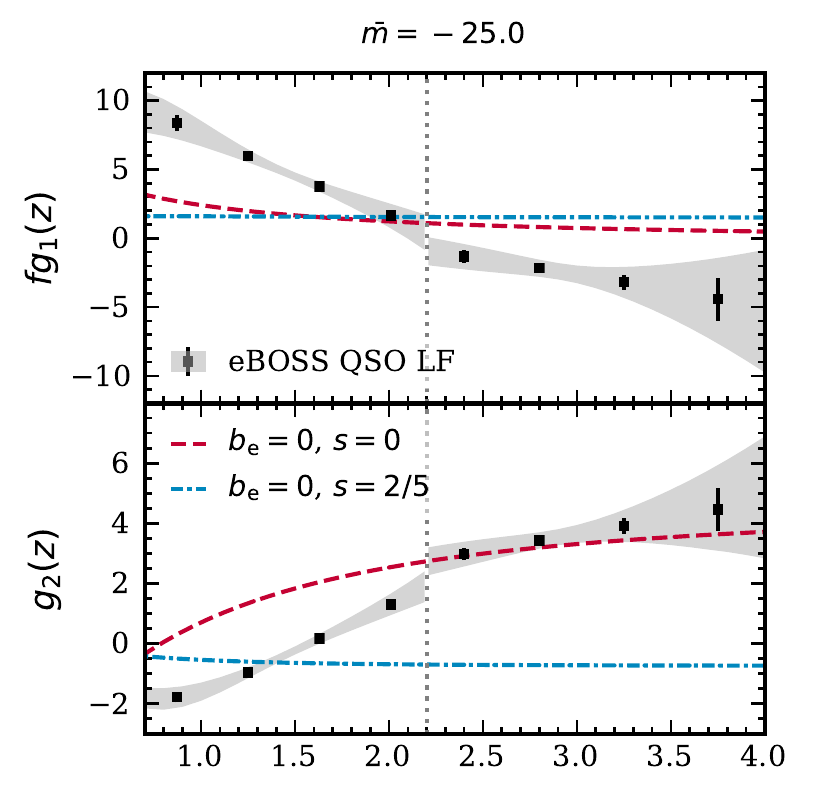} \\[-3pt]
    \caption{Derived measurements of the relativistic correction \change{functions $f g_1(z)$ and $g_2(z)$} (see equation~\ref{eq:relativistic contributions}) at the absolute magnitude threshold $\mbar = -25$ from the best-fitting eBOSS QSO LF in this work (see Table~\ref{tab:PLE model fits}). The data points with error bars are measurements within the \SI{68}{\percent}~credible interval for the eBOSS QSO redshift bins, and the shaded region shows the \SI{95}{\percent} credible interval. \change{For comparison, the dashed red lines show the results with $(b_\evol, s) = (0, 0)$ and the dash-dotted blue lines with $(b_\evol, s) = (0, 2/5)$.} The vertical dotted line marks the pivot redshift $z_\pivot = 2.2$.}
    \label{fig:relativistic corrections}
\end{figure}

In section~\ref{sec:relativistic clustering}, \change{we have shown that relativistic corrections of $\order{\cH/k}$~and~$\order*{\cH^2/k^2}$ to the galaxy overdensity field at different redshifts and scales are modulated by the functions $f g_1(z)$~and~$g_2(z)$ respectively, which can be constrained from the relativistic bias measurements obtained above under the Planck15 cosmology. In Fig.~\ref{fig:relativistic corrections}, we show the derived bounds on $f g_1(z)$~and~$g_2(z)$ within the \SI{95}{\percent}~credible interval and their measurements in eBOSS QSO redshift bins with \SI{68}{\percent}~level uncertainties. The discontinuities in the derived $g_1(z)$~and~$g_2(z)$ have the same origin as those discussed previously. The values and uncertainties of $g_1(z)$~and~$g_2(z)$} are both dominated by contributions from evolution bias $b_\evol$, which can be an order of magnitude larger than $s(z)$ as shown in Fig.~\ref{fig:relativistic biases}. \change{To assess the importance of $b_\evol$~and~$s$, we have also shown in Fig.~\ref{fig:relativistic corrections} two interesting fiducial cases: $(b_\evol, s) = (0, 0)$, i.e. no account of the redshift evolution and luminosity dependence of the tracer number density; $(b_\evol, s) = (0, 2/5)$, i.e. the comoving number density is constant and the common factor~$(2 - 5s)$ vanishes in relativistic corrections, corresponding to the so-called `diffuse background' scenario where the effects of lensing magnification and volume distortions partly cancel~\citep{Jeong_2012}. Comparisons with these cases demonstrate that evolution bias $b_\evol$ drives relativistic corrections at both low and high redshifts; unless $(2 - 5s)$ vanishes, terms containing the $(\cH \chi)^{-1}$~factor are also important and increasingly so at lower redshift (especially beyond the redshift range shown in the figures towards $z = 0$). This highlights the importance of including accurate models of both $b_\evol$~and~$s$ in relativistic corrections to galaxy clustering.}

Having propagated quasar luminosity function measurements through to constraints \change{on relativistic corrections}, we shall investigate in the following section how \change{they modify} the quasar clustering power spectrum multipoles on large scales.

\section{\texorpdfstring{\textls{Scale-dependent Modification to Power Spectrum Multipoles}}{Scale-dependent Modification to Power Spectrum Multipoles}}
\label{sec:scale-dependent modifications}

In the presence of local primordial non-Gaussianity~$\fNL$, the linear tracer bias~$b_1(z)$ receives a scale-dependent modification
    \begin{equation}
        \upDelta b(k, z) = 3 \fNL (b_1 - p) \frac{1.27 \delta_\textrm{c} \Omega_{\matter,0} H_0^2}{c^2 k^2 T(k) D(z)} \,,
    \end{equation}
where $p$ is a parameter that depends on the tracer sample (here we adopt $p = 1.6$ for quasars), $H_0$ is the Hubble parameter at present, $c$ is the speed of light, and $T(k)$ is the matter transfer function \citep{Slosar_2008}. The numerical factor~$1.27$ arises as we normal{\is}e the linear growth factor~$D(z)$ to unity at present. As $k \rightarrow 0$, $T(k) \rightarrow 1$ and $\upDelta b \propto k^{-2}$, so the signal of~$\fNL$ is enhanced on large scales.

Under the plane-parallel approximation, \citet{Kaiser_1987} showed the anisotropic clustering power spectrum in redshift space is
    \begin{equation}
        P^\textrm{K}(k, \mu) = \qty\big(b + f \mu^2)^2 P_\matter(k)
    \end{equation}
on large scales, where $P_\matter$ is the linear matter power spectrum. By considering the Legendre multipoles with respect to the angle variable $\mu$, $P^\textrm{K}(k, \mu)$ is equivalent to the combination of the monopole
\begin{subequations}
    \begin{equation}
        P^\textrm{K}_0(k) = \qty(b^2 + \frac{2}{3} b f + \frac{1}{5} f^2) P_\matter(k) \,,
    \end{equation}
the quadrupole
    \begin{equation}
        P^\textrm{K}_2(k) = \qty(\frac{4}{3} b f + \frac{4}{7} f^2) P_\matter(k)
    \end{equation}
\end{subequations}
and the hexadecapole~$P^\textrm{K}_4(k)$ which we neglect as it does not depend on the tracer bias. Note that here the total bias~$b$ now includes both $b_1$ and the scale-dependent modification~$\upDelta b \propto k^{-2}$, which changes the standard Kaiser multipoles~$P^\textrm{K}_\ell$ with only the scale-independent bias~$b_1$ by
\begin{subequations}
    \label{eq:non-Gaussianity multipole modifications}
    \begin{align}
        & \upDelta P_0(k) = \qty[\qty(2 b_1 + \frac{2}{3} f) \upDelta b + {\upDelta b}^2] P_\matter(k) \,, \\
        & \upDelta P_2(k) = \frac{4}{3} \upDelta b f P_\matter(k) \,.
    \end{align}
\end{subequations}
In contrast to the quadrupole which only receives a modification proportional to $k^{-2}$, the monopole receives modifications proportional to both $k^{-2}$~and~$k^{-4}$ when $\fNL \neq 0$.

In section~\ref{sec:relativistic clustering}, we have shown that relativistic corrections similarly leave a scale-dependent signature. By considering the two-point function of expression~\eqref{eq:relativistic corrections}, we see that relativistic corrections only change the Kaiser monopole and quadrupole by
\change{%
\begin{subequations}
    \label{eq:relativistic multipole modifications}
    \begin{multline}
        \upDelta P_0(k) = \bigg[\bigg(2 b_1 g_2 + \frac{2}{3} f g_2 + \frac{1}{3} f^2 g_1^2\bigg) \\ \times \qty(\frac{\cH}{k})^2 + g_2^2 \qty(\frac{\cH}{k})^4\bigg] P_\matter(k) \,, \qquad
    \end{multline}
    \begin{align}
        & \upDelta P_2(k) = \frac{2}{3} \qty(2 f g_2 + f^2 g_1^2) \qty(\frac{\cH}{k})^2 P_\matter(k) \,.
    \end{align}
\end{subequations}
By comparing equations~\eqref{eq:non-Gaussianity multipole modifications}~and~\eqref{eq:relativistic multipole modifications}, it is evident that relativistic corrections can mimic the effect of~$\fNL$ in both the monopole and quadrupole of the clustering power spectrum on large scales; the extent to which relativistic corrections can wash out the $\fNL$~signal depends on the precise amplitudes of $g_1(z)$~and~$g_2(z)$.}

Since we have obtained \change{constraints on $g_1(z)$~and~$g_2(z)$} in the previous section, we can make a concrete comparison between the power spectrum multipole modifications due to $\fNL$ and \change{relativistic corrections}. To this end, we consider a fiducial value~$\fNL = 1$ at which level different classes of inflation models can be distinguished~\citep{Maldacena_2003,Alvarez_2014}. As the fiducial case, we continue to adopt the Planck15 cosmology, the absolute magnitude threshold $\mbar = -25$ for the quasar sample\change{, and $b_1(z) = \flatfrac{1.2}{D(z)}$ as the baseline assumption for DESI~\citep[DESI Collaboration,][]{DESI_2016}, which is also used in the UMF prediction (see section~\ref{subsec:relativistic bias constraints})}.

\begin{figure*}
    \centering
    \includegraphics[width=0.49\linewidth]{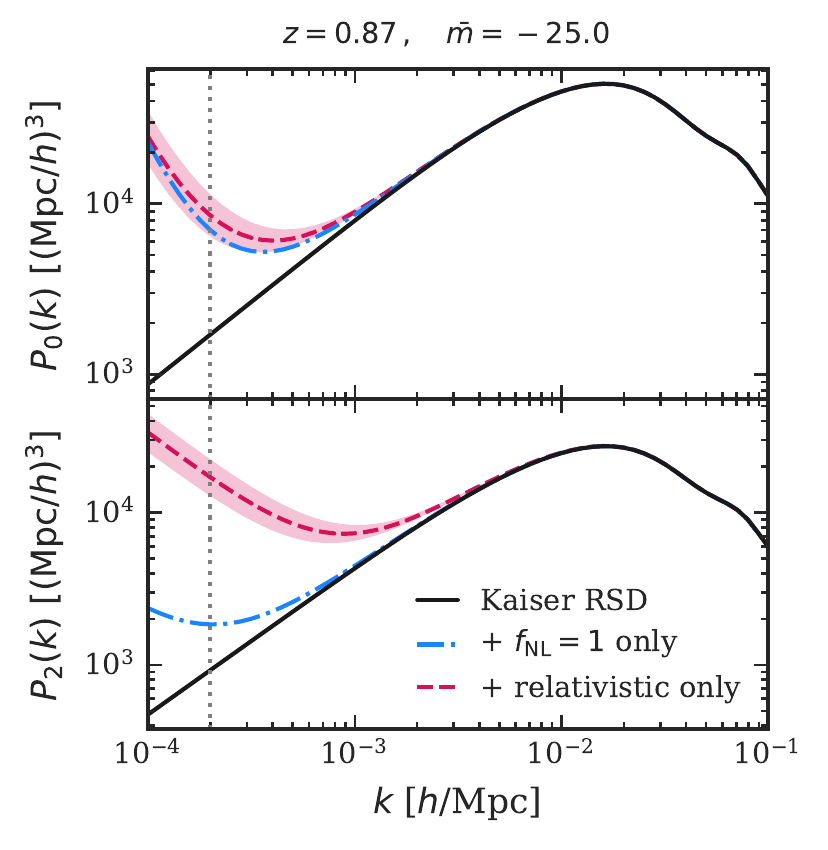}
    \hfill
    \includegraphics[width=0.49\linewidth]{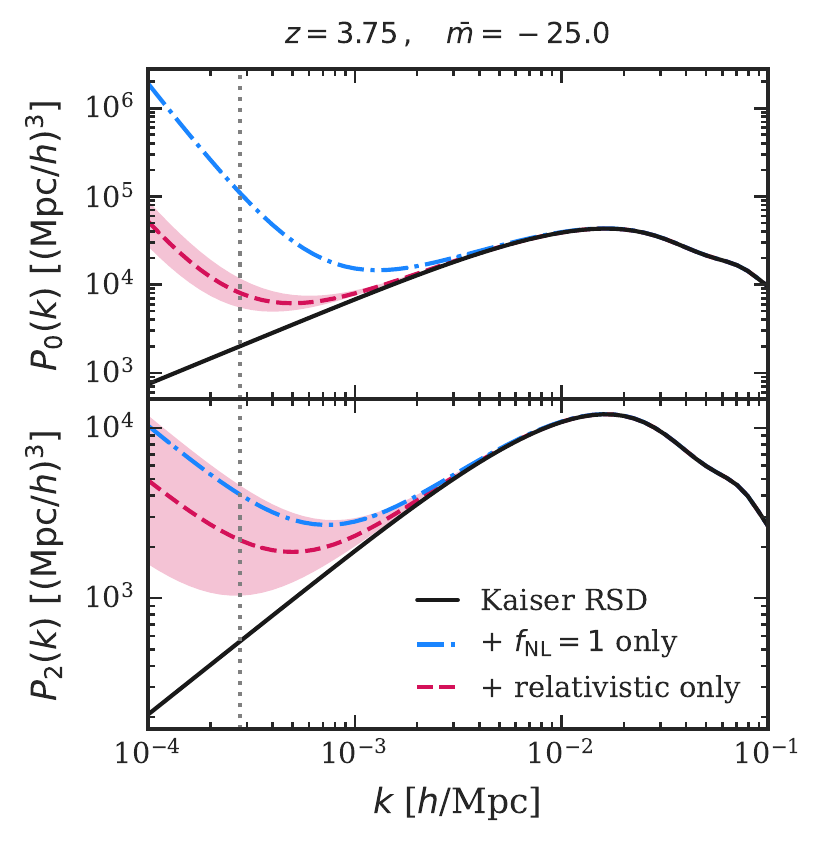} \\[-3pt]
    \caption{Large-scale quasar clustering power spectrum monopole~$P_0(k)$ and quadrupole~$P_2(k)$ at redshift~$z = 0.87$ (\emph{left-hand column}) and $z=3.75$ (\emph{right-hand column}) with magnitude threshold~$\mbar = -25$. The Kaiser RSD model is shown by the solid black lines. The effect of relativistic corrections without local non-Gaussianity~$\fNL$ is shown by the dashed red lines with the shaded red regions showing the \SI{95}{\percent}~credible interval derived from the best-fitting eBOSS QSO LF in this work (see Table~\ref{tab:PLE model fits}). The effect of~$\fNL = 1$ (fiducial value) without relativistic corrections is shown by the dash-dotted blue lines. The vertical dotted lines mark the horizon scale~$k = \cH$.}
    \label{fig:clustering multipoles}
\end{figure*}

In Fig.~\ref{fig:clustering multipoles}, we show the power spectrum multipoles for~$k \in [10^{-4}, 10^{-1}] \, \si{\h\per\mega\parsec}$ in the presence of these modifications at two redshifts, $z = 0.87$~and~$3.75$, which we recall are respectively the lowest and highest eBOSS QSO redshift bins. At the lower redshift, relativistic effects dominate over the $\fNL$~signal and obscure the PNG signature. \change{At the higher redshift, although the relativistic modification is almost comparable to the $\fNL$~effect in the quadrupole, the $\fNL$~signal is larger in the monopole.} This offers a hint that, at least for the quasar sample, pushing the upper redshift range may \change{help mitigate some potential contamination of the $\fNL$~signal from part of the relativistic corrections}; \change{however, we caution that lensing convergence and non-local potential terms have not been included in our analysis, and these integrated contributions might hamper the detection of PNG again at higher redshifts~\citep[see e.g.][]{Namikawa_2011,Raccanelli_2016b,Lorenz_2018}.}

\begin{figure*}
    \centering
    \includegraphics[width=0.49\linewidth]{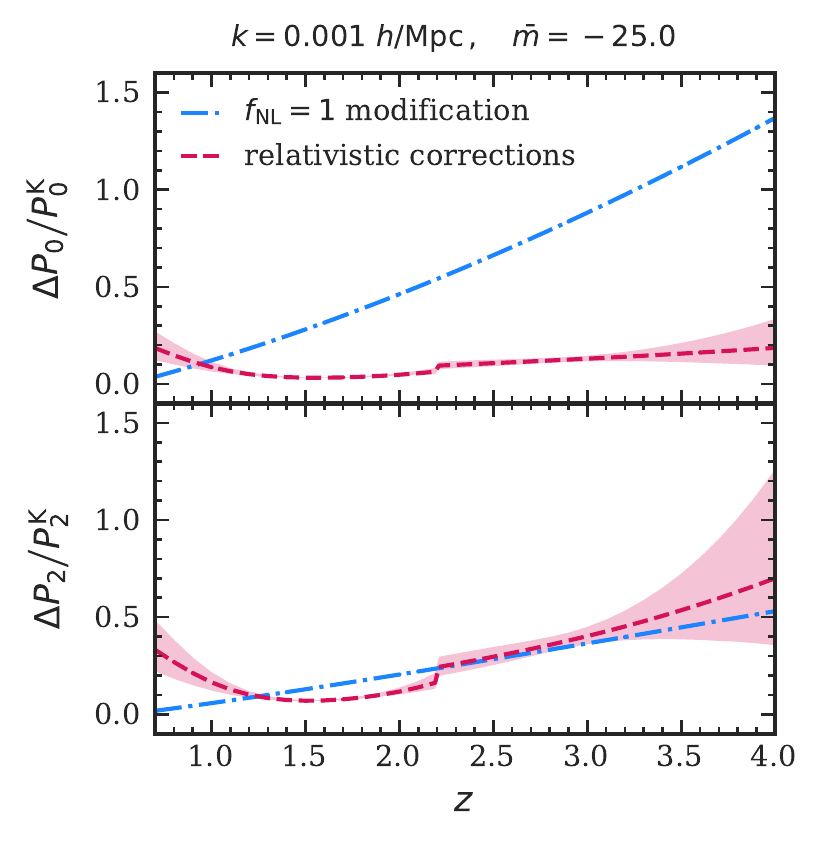}
    \hfill
    \includegraphics[width=0.49\linewidth]{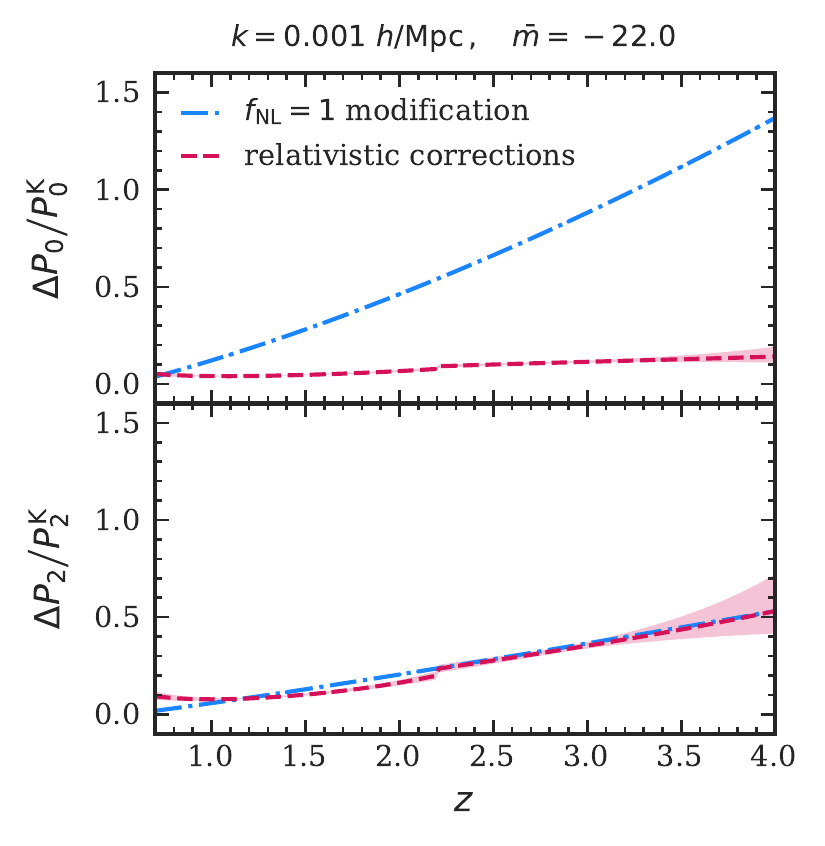} \\[-3pt]
    \caption{Scale-dependent modifications~$\Delta P_{\ell}$ to the quasar clustering power spectrum monopole and quadrupole as a fraction of the Kaiser RSD model~$P^\textrm{K}_{\ell}$ at wave number~$k = 0.001 \si{\h\per\mega\parsec}$ with absolute magnitude threshold~$\mbar = -25$ (\emph{left-hand column}) and $\mbar = -22$ (\emph{right-hand column}). Relativistic corrections without local non-Gaussianity~$\fNL$ are shown by the dashed red lines with the shaded red regions showing the \SI{95}{\percent}~credible interval derived from the best-fitting eBOSS QSO LF in this work (see Table~\ref{tab:PLE model fits}). Modifications due to $\fNL = 1$ (fiducial value) without relativistic corrections are shown by the dash-dotted blue lines.}
    \label{fig:scale-dependent modifications}
\end{figure*}

\change{To have a broader view of how the relative amplitudes of relativistic and PNG modifications evolve with redshift, in Fig.~\ref{fig:scale-dependent modifications} we compare the change in power spectrum multipoles, $\upDelta P_\ell$,} as a fraction of the standard Kaiser prediction~$P^\textrm{K}_\ell$ across the eBOSS QSO redshift range~$0.7 < z < 4$ at a fixed wave number~$k = 10^{-3} \si{\h\per\mega\parsec}$ -- this is close to the largest scale which DESI and \textit{Euclid} may access \change{(DESI Collaboration, \citealt{DESI_2016}; Euclid Consortium, \citealt{Euclid_2011})}. In addition to our fiducial absolute magnitude threshold~$\mbar = -25$, we also consider a less conservative cut at~$\mbar = -22$, which is the limiting magnitude of the lowest eBOSS QSO redshift bin. \change{The discontinuities seen in Fig.~\ref{fig:scale-dependent modifications} have the same origin as those found in relativistic bias constraints in the previous section. We summar{\is}e the key findings from the figure as follows:%
\begin{enumerate}
    \item For both monopole and quadrupole, relativistic corrections dominate over the effect of~$\fNL$ at low redshifts~$z \lesssim 1$, and values of~${\upDelta P_\ell}$ due to relativistic effects and $\fNL$ reach parity at some intermediate redshift below~$z = 1.5$. The dominance of relativistic effects at lower redshifts is mainly driven by the large evolution bias~$b_\evol$ (see Fig.~\ref{fig:relativistic biases}) and the geometric factor~$(\cH \chi)^{-1}$ when~$s \neq 2/5$. If $b_\evol = 0$~and~$s = 2/5$, relativistic effects will be much smaller than the $\fNL$~signal overall;
    \item Although relativistic corrections are comparable to the $\fNL$~effect at most redshifts in the quadrupole, the $\fNL$~signal is stronger at higher redshifts in the monopole, mainly because of the redshift evolution of linear tracer bias~$b_1(z)$ and the fact that $\upDelta P_0$ due to $\fNL$ contains contributions proportional to~$b_1^2$. If the tracer bias is constant, say $b_1 = 2$, then at higher redshifts, relativistic effects with $b_\evol \neq 0$~and~$s \neq 2/5$ will wash out the $\fNL$~signal, and can slightly reduce the $\fNL$~signal even with $b_\evol = 0$~and~$s = 2/5$;
    \item Raising the absolute magnitude threshold tends to reduce the relativistic corrections at all redshifts: we have checked that both evolution and magnification bias decreases in absolute values with increasing magnitude threshold, suggesting that future surveys with sensitivity to detect more fainter objects could also help with constructing tracer samples with subdued relativistic effects.
\end{enumerate}}

\change{It is worth mentioning that in the limit~$k \rightarrow 0$, \citet{Grimm_2020} have recently shown that the full relativistic effects actually vanish as a consequence of the equivalence principle, and thus they do \emph{not} contaminate the PNG signature. The apparent divergence in~$\upDelta P_\ell$ as~$k \rightarrow 0$ in equation~\eqref{eq:relativistic multipole modifications} is due to the exclusion of non-local contributions as well as contributions at the observer position. However, for finite clustering scales accessible to galaxy surveys, these relativistic effects do exist and thus should be taken into account in PNG analysis.}

\change{For cosmological parameter inference} from clustering measurements made on very large scales, the control of large-scale systematics should closely accompany the inclusion of relativistic corrections. For instance, the plane-parallel limit for power spectrum multipoles has been taken to simplify arguments in this work, but wide-angle effects due to variation of the line of sight have been shown to be critical on very large scales \citep{Szalay_1998,Szapudi_2004,Papai_2008,Yoo_2015}. Therefore in a practical analysis, wide-angle corrections need to be included perturbatively \citep{Castorina_2018,Beutler_2019}, or a spherical Fourier analysis could prove advantageous \citep{Fisher_1995,Heavens_1995,Yoo_2013,Wang_2020b}. Meanwhile, we have only considered quasars as a single tracer for detecting relativistic effects and the PNG signature in this work; to extract maximal information from future LSS probes, a multitracer approach can enhance the signal-to-noise ratio and thus prove more beneficial \citep{McDonald_2009,Seljak_2009}.

\section{\texorpdfstring{\textls{Conclusion}}{Conclusion}}
\label{sec:conclusions}

Motivated by recent studies of relativistic effects in LSS observations and the prospect of constraining PNG through galaxy redshift surveys to the level of precision competitive with CMB experiments in the near future, we have sought to quantify relativistic corrections to clustering statistics on very large scales, with quasars as a concrete example. These corrections do not only depend on the cosmological expansion history, but also on the redshift evolution of the underlying quasar number density and its sensitivity to the luminosity threshold, which are parametr{\is}ed by evolution bias $b_\evol$ and magnification bias $s$. To this end, we have refitted the eBOSS QSO luminosity function and derived measurements on both $b_\evol$ and $s$, before propagating their constraints to relativistic corrections to the power spectrum multipoles. Our assessment of the impact of relativistic effects on the $\fNL$ signature affirms the results of previous works mentioned in section~\ref{sec:introduction}, but this agreement is reached after a more realistic treatment for evolution and magnification bias contributions, in particular their uncertainties.

We have found that relativistic corrections can indeed be mistaken for $\fNL$-induced scale-dependent bias modifications, especially at low redshifts and in the power spectrum quadrupole. By using tracer samples at higher redshifts or with a fainter luminosity threshold, relativistic effects can be reduced to some extent. We have also found that, at least for the quasar population, the impact of evolution bias $b_\evol$ and its uncertainties on clustering statistics is greater than that of magnification bias $s$. However, the latter also appears in lensing contributions to the observed galaxy overdensity field, which we have neglected in this work \change{along with other integrated terms involving the gravitational potential; these contributions can be important especially at higher redshifts, and are best studied in future works with alternative statistics such as the angular or spherical power spectrum.}

For future clustering measurements of the DESI quasar sample with apparent magnitude limit similar to the one considered in this work \citep[DESI Collaboration,][]{DESI_2016}, relativistic corrections can be an order-of-magnitude larger than the modifications induced by $\fNL \simeq 1$ on scales $k \sim \SI{e-3}{\h\per\mega\parsec}$ at lower redshifts $z \lesssim 1$; \change{at higher redshifts $z \gtrsim 2$, relativistic corrections remain comparable to or larger than the $\fNL \simeq 1$ modification in the power spectrum quadrupole for absolute magnitude threshold up to $\mbar = -22$ at least. We have seen in section~\ref{sec:relativistic constraints} how potential systematics in the quasar luminosity function can affect the relativistic bias parameters, and therefore the accurate determination of tracer luminosity functions is crucial to constraining relativistic corrections and local primordial non-Gaussianity. We suggest that forward modelling from the tracer luminosity function to relativistic corrections should be fully included in future cosmological analysis.} For this purpose, we have made the code implemented in this work publicly available as a Python package, \textsc{HorizonGRound}\footnote{\href{https://github.com/MikeSWang/HorizonGRound/}{\texttt{github.com/MikeSWang/HorizonGRound}}}.

\section*{\textls{Acknowledgements}}

\change{The authors would like to thank the anonymous referee for helpful feedback.} The authors would also like to thank Obinna Umeh and Christophe Y\`{e}che for helpful discussions, and Nathalie Palanque-Delabrouille and Christophe Magneville for insights on the luminosity function fitting procedure. MSW would also like to thank Minas Karamanis for help with the MCMC sampling code \textsc{zeus} which is publicly available.

MSW is supported by the University of Portsmouth Student Bursary. FB is a Royal Society University Research Fellow. DB is supported by the UK STFC grant ST/S000550/1.

Numerical computations are performed on the Sciama High Performance Computing (HPC) cluster which is supported by the Institute of Cosmology and Gravitation (ICG), the South East Physics Network (SEPnet) and the University of Portsmouth. This work has made use of publicly available Python packages \textsc{astropy} (Astropy Collaboration, \citealt{astropy:2013}, \citealt{astropy:2018}) and \textsc{nbodykit}~\citep{Hand_2018}.

\section*{\textls{Data Availability}}

\change{%
The data underlying this article, including the code used to derive further data, are available at \url{https://github.com/MikeSWang/HorizonGRound}. All data sets have been derived from the following source in the public domain: A\&A 587, A41 (2016) (DOI: \href{https://dx.doi.org/10.1051/0004-6361/201527392}{10.1051/0004-6361/201527392}).
}

{
    \raggedright
    \hypersetup{urlcolor=MNRASPurple}
    \bibliographystyle{mnras}
    \bibliography{references}
}

\bsp

\label{lastpage}
\end{document}